\documentclass[a4paper,aps,superscriptaddress,floatfix,nofootinbib,twocolumn,bibtex]{revtex4}

\usepackage{bbold}
\usepackage{bbm}
\usepackage[pdftex]{graphicx}
\usepackage{latexsym,amsmath,verbatim}
\usepackage{color}
\usepackage{rotating}
\usepackage{verbatim}
\usepackage{multirow}
\usepackage[english]{babel}
\usepackage{comment}

\usepackage{amsmath,amssymb}
\usepackage{romannum}

\begin{document}

\title{Observability transition in multiplex networks}

\author{Saeed Osat}
\affiliation{Molecular Simulation Laboratory, Department of Physics, Faculty of Basic Sciences, Azarbaijan Shahid Madani University, Tabriz 53714-161, Iran}

\author{Filippo Radicchi}
\affiliation{Center for Complex Networks and Systems Research, School
  of Informatics and Computing, Indiana University, Bloomington,
  Indiana 47408, USA}
\email{filiradi@indiana.edu}

\begin{abstract}
 We extend the observability model
 to multiplex networks. We present mathematical frameworks, valid
 under the treelike ansatz, able to
describe the emergence of the macroscopic cluster
of mutually observable nodes
in both synthetic and real-world multiplex networks. 
We show that the observability transition in synthetic multiplex networks
is discontinuous. In real-world multiplex networks instead,
edge overlap among layers is responsible for the
disappearance of any sign of abruptness in the emergence of the
the macroscopic
 cluster
of mutually observable nodes.
\end{abstract}

\maketitle

\section{Introduction}

Complex systems where elementary units 
have different types
of interactions can be conveniently modelled as
multiplex networks~\cite{boccaletti2014structure,
  kivela2014multilayer, lee2015towards}.
This is a very generic representation, where
elements of a system are organized
in multiple network layers, 
each standing for a specific color or flavour of interaction.
Systems that can be represented in this way
are abundant in  the real world. Examples include, among others, 
social networks sharing the same
actors~\cite{szell2010multirelational, mucha2010community}, and
multimodal transportation graphs sharing 
common geographical
locations~\cite{barthelemy2011spatial,cardillo2013emergence}.

Several  analyses
of multiplex networks have been performed 
recently~\cite{boccaletti2014structure,
  kivela2014multilayer, lee2015towards}. A common 
result, shared by the vast majority of these studies,
is that a processes defined on a multiplex network
is characterized by features radically different from those
observed for the same type of process when this is applied to an isolated
network. Examples regard dynamical 
processes taking place on multiplex networks, 
such as diffusion~\cite{PhysRevLett.110.028701,
  de2014navigability}, 
epidemic spreading~\cite{PhysRevE.85.066109, PhysRevE.86.026106,
  PhysRevLett.111.128701, de2016physics}, 
synchronization~\cite{delGenioe1601679}, and 
controllability~\cite{PhysRevE.94.032316}.
Examples include also 
structural processes as those typically framed
in terms of percolation models.
For instance in their seminal paper, 
Buldyrev {\it et al.}  considered a
site-percolation model aimed at
understanding the role of
interdependencies among two network
layers~\cite{buldyrev2010catastrophic}.
The macroscopic cluster of mutually connected nodes in a multiplex
network emerges
discontinuously, at odds with
what instead observed for the same process on an isolated network
where the percolation transition is always continuous. 
A large number of subsequent studies have 
then analyzed in detail the features of
the percolation transition in multiplex 
networks~\cite{parshani2010interdependent, parshani2011inter, baxter2012avalanche, PhysRevE.89.012808, son2012percolation, min2014network, PhysRevE.91.012804, bianconi2014multiple, radicchi2013abrupt, radicchi2015percolation, cellai2016multiplex, PhysRevE.94.060301}.
Several variants of the percolation
model have  been also considered, including
redundant site percolation~\cite{radicchi2016redundant},
k-core percolation~\cite{PhysRevE.90.032816}, weak
percolation~\cite{PhysRevE.89.042801}, and
bond percolation~\cite{PhysRevX.6.021002}.

In this paper, we focus our attention on an additional 
variant of the percolation model 
usually named as the observability 
model~\cite{PhysRevLett.109.258701, PhysRevE.94.030301}.
In isolated networks,
the model finds its motivation in the study of 
some dynamical processes where the state of 
the system can be determined by
monitoring or dominating  
the states of a limited 
number of nodes in the network~\cite{liu2011controllability}. 
Examples include, among others, real-time monitoring
of power-grid networks~\cite{PhysRevLett.109.258701} and
mobile ad-hoc networks~\cite{wu1999calculating}.
The observability model has been considered in 
synthetic models~\cite{PhysRevLett.109.258701} and real-world
topologies~\cite{PhysRevE.94.030301}. In the simplest version 
of the model, every node in the network can host 
an observer with probability $\phi$. Placing an observer on one node can 
make the node itself and all its nearest neighbors observable.
Nodes in the network can therefore assume three different
states: (i) directly observable, if hosting an observer; (ii) indirectly
observable, if being the first neighbor of an observer; (iii) or not
observable, otherwise. 
Observable, either directly 
or indirectly, nearest-neighbor nodes form clusters
of connected observable nodes. As in the case of percolation, the 
question of  interest in network observability
is understanding the macroscopic formation of observable clusters
in the network on the basis of microscopic changes
in the state of its individual nodes. 
In synthetic infinite graphs, the macroscopic
cluster of observable nodes obeys a continuous phase transition 
as a function of the probability
$\phi$, and the critical threshold is generally very small~\cite{PhysRevLett.109.258701}.
In most real-world networks also, the largest cluster grows smoothly,
and the transition point is generally very close to zero~\cite{PhysRevE.94.030301}.

In our extension of the observability model to multiplex networks, we
focus our attention on the emergence of clusters
of mutually observable nodes. The definition of these clusters is a
straightforward
combination of the notions of clusters of observable nodes
in isolated networks~\cite{PhysRevLett.109.258701} and clusters 
of mutually connected nodes in
multiplex networks~\cite{buldyrev2010catastrophic}.
The extension of the observability model to multiplex networks
finds its rationale in any situation where the goal is 
monitoring
or controlling the dynamics of a system structured in multiple layers of 
interactions. A genuine example could be tracking the spread of information
on two coupled social media, such as Facebook and Twitter.
Actors are shared by the two social networks. 
Observing an individual means getting full access to her/his
accounts thus being able to read the content
of the messages that the individual, and 
her/his friends, are posting on the two platforms. 
Clusters of mutually observable nodes in this sense correspond to
connected portions of the systems where 
the content of information that is exchanged by people 
within the cluster can be completely monitored. 

The paper is organized as follows. In section~\ref{sec:intro}, we
define the observability model in multiplex networks. In
section~\ref{sec:random},
we study the model in ensembles of multiplex networks
whose layers are generated according to the configuration model.
In section~\ref{sec:real}, we introduce a message-passing method
to deal with the observability model in multiplex networks with specified
topology, such as real-world multiplex networks. The framework
described in this section is based on the assumption that the
number of edges shared by the different layers of the multiplex
is negligible. The message-passing method valid for multiplex networks
in presence of edge overlap is much more cumbersome, and, for this
reason,
presented only in the Supplemental Material. Finally, in
section~\ref{sec:conclusions}, we summarize the 
main findings of the paper.

\section{Observability model in multiplex networks}
\label{sec:intro}
We consider a multiplex network composed of $N$ nodes structured in two 
layers, namely $\alpha$ and $\beta$.
Node labels take integer values from $1$ to $N$
in both layers. Nodes with identical labels
correspond to copies (or replicas) of the same individual or unit 
in the two layers. The observability 
model we consider here is a natural extension to multiplex
networks
of the same model already considered 
on isolated networks~\cite{PhysRevLett.109.258701, PhysRevE.94.030301}.
Observers or sensors are placed at random with probability
$\phi$ on every node in the system. If an observer is placed on a node
$i$,
the node $i$ is directly observable in both layers. A node $i$, that is not
directly
observable, but is attached to at least one directly observable node $j$
in layer $\alpha$ and at least one directly observable node $k$ in
layer $\beta$, with $j$ not necessarily equal to $k$, is indirectly observable.
We focus our attention on mutually observable clusters of nodes.
The way these clusters are defined is identical to the one
in which clusters of mutually connected nodes are defined in
site percolation~\cite{buldyrev2010catastrophic}. 
The only difference comes from the fact that 
a node in order to be ``occupied'' can be either in the directly or indirectly
observable state. Note that mutually connected and mutually observable clusters
coincide for $\phi = 1$.
 In particular, a cluster of mutually observable nodes
is defined in a recursive manner and is composed by all the nodes 
that are connected by at least by one path (internal to the cluster
itself) 
in both layers. Our model can be 
seen as a depth-one percolation model on a multiplex network.
As we are interested in understanding the extent of the system
that can be monitored by placing random points of
observation, the focus of our analysis is centered on
quantifying how the size of
the Largest Mutually Observable Cluster (LMOC) 
changes as a function 
of the microscopic probability $\phi$ of nodes to be directly
observable. 
To generate the mathematical framework necessary
to study such a model, we make use of a simple self-consistent
condition: an observable node belongs to the
LMOC if it has, in both layers, at least one 
neighbor that is part of the LMOC. Such a condition can be expressed 
in terms of elementary conditional probabilities defined for the edges
and the nodes
of the multiplex network. In the following, we discuss the details of
how the mathematical framework can be solved exactly
under two conditions: (i) absence of link overlap among layers, and
(ii) validity of the locally treelike ansatz. In 
the Supplemental Material, we report 
the full mathematical framework valid 
when condition (i) is removed.

\section{Observability transition in random multiplex networks}
\label{sec:random}
Let us consider the case of a multiplex composed of two 
network layers generated independently according to the configuration model~\cite{molloy1995critical}.
If the layers are sufficiently sparse, the fact that the layers
are generated independently allow us to consider the overlap among layers (i.e., the simultaneous existence of the
same edge in both layers happens with vanishing probability) negligible.
The only inputs required to study such a system are the degree 
distributions
of the two layers, namely $P^{[\alpha]}(k)$ and $P^{[\beta]}(k)$.

The analytic treatment for this special types of networks
is similar to one presented in Ref.~\cite{PhysRevLett.109.258701}
for isolated networks. To generate a self-consistent set of equations 
able to describe how the LMOC changes as a function of $\phi$ in this special 
type of multiplex networks, we define the following conditional probabilities valid 
for a randomly selected edge in layer $\alpha$:

\begin{enumerate}

\item $u^{[\alpha]}$, that is the probability that one of the nodes at the end of the edge
is in the LMOC given that the other node at the end of the
  edge  is directly observable.

\item $v^{[\alpha]}$, that is the probability that one of the nodes at the end of the edge
is in the LMOC given that the node at the other end of the edge is not directly observable.

\item $z^{[\alpha]}$, that is the probability that one of the nodes at the end of the edge
is in the LMOC given that both nodes at the end of the edge are not directly observable.

\end{enumerate}
The same exact definitions are valid for the conditional probabilities 
$u^{[\beta]}, v^{[\beta]},$ and $z^{[\beta]}$ for layer $\beta$.

As mentioned above,
a generic node is part of the LMOC if observable, either directly
or indirectly, and attached to at least one
other node that is part of the LMOC. The way this happens
depends however on the state of the node.
We make therefore a distinction between (i) a directly observable 
node and (ii) a not directly observable node.

Let us consider first case (i). If one of the nodes at the end
of a generic edge is directly observable, the probability 
that the node is not connected to
the LMOC is $\phi (1 - u^{[\alpha]})$ if the other node is
directly observable, or $(1-\phi)(1-v^{[\alpha]})$ if the
other node is not directly observable. If 
the node we are considering
has degree $k$, then then probability that the node is
connected to the LMOC in layer $\alpha$ is
\[
q^{[\alpha]}_k = 1 - \sum_{m=0}^{k} {k \choose m} [\phi
(1-u^{[\alpha]})]^m [(1-\phi)(1-v^{[\alpha]})]^{k-m} \; .
\]
The sum on the r.h.s. of the equation above quantifies the probability 
for the none of the neighbors of a node with degree $k$ is part of the LMOC.
This probability is then discounted from $1$ to
estimate the probability that at least one neighbor of the node
is part of the LMOC.
If we consider all nodes, we have that
\[
q^{[\alpha]} = \sum_k P^{[\alpha]}(k) q^{[\alpha]}_k = 1 - G_0^{[\alpha]} \left(1
- \phi u^{[\alpha]} - (1-\phi) v^{[\alpha]} \right) \;,
\]
where $G_0^{[\alpha]}(x) = \sum_k P^{[\alpha]}(k) x^{k}$ is the
generating function of the degree distribution for layer $\alpha$.
We can repeat the same exact  arguments for layer $\beta$.
In particular, the probability that a generic node 
is part of the LMOC is given by the product that the node is attached
to the LMOC simultaneously in both layers, that is
\begin{equation}
q = \phi \, q^{[\alpha]}  \, q^{[\beta]} \; ,
\label{eq:q}
\end{equation}
where
the extra factor $\phi$ comes from the fact that we are
considering the case of a node that is directly observable.

For case (ii), we can proceed in a similar way as above.
If our node is not directly observable, and we select one of its edges
at random, then
the probability that this
node is not connected to
the LMOC is $\phi (1 - u^{[\alpha]})$ if the other node at the end of
the edge is
directly observable, or $(1-\phi)(1-z^{[\alpha]})$ if the
other node is not directly observable.
If  the node we are considering
has degree $k$, the probability that this node is connected to the LMOC in layer $\alpha$ is
is
\[
\begin{array}{ll}
r^{[\alpha]}_k = & 1-  (1- \phi)^k + 
\\
& - \sum_{m=1}^{k} {k \choose m} [\phi
(1-u^{[\alpha]})]^m [(1-\phi)(1-z^{[\alpha]})]^{k-m} 
\end{array} \; .
\]
The term $(1-\phi)^k$ comes from the fact that
if none of the nodes at the end of the $k$ edges departing
from our node are directly observable, then necessarily our node will be
not
indirectly observable and thus surely out of any cluster.
The remaining part of the r.h.s. instead quantifies the probability that
none of the neighbors of our node is part of the LMOC, assuming that
at least one of them is directly observable. Both these probabilities
are discounted from $1$ to compute the probability
that our indirectly observable node is part of the LMOC.
If we consider all nodes, we have that
\[
\begin{array}{l}
r^{[\alpha]} =  \sum_k P^{[\alpha]}(k) r^{[\alpha]}_k =
 \\
1 - G_0^{[\alpha]} \left(1
- \phi u^{[\alpha]} - (1-\phi) z^{[\alpha]} \right)  +
\\ +  G_0^{[\alpha]}
  \left( (1- \phi) (1 - z^{[\alpha]}) \right) - G_0^{[\alpha]}
  \left( 1- \phi \right)
\end{array}
\; .
\]
For layer $\beta$, the arguments are identical.
The probability that a generic node that is not directly observable
is part of the LMOC is given by the product that the node is attached
to the LMOC simultaneously in both layers, that is
\begin{equation}
r = (1 - \phi)  \, r^{[\alpha]} \, r^{[\beta]} \; ,
\label{eq:r}
\end{equation}
where
the extra factor $1 - \phi$ comes from the fact that we are
considering the case of a node that is not directly observable.
Combining cases (i) and (ii) together, we can finally write
\begin{equation}
P_\infty =  q + r \; ,
\label{eq:lmoc}
\end{equation}
for the average size of the LMOC. To compute
Eq.~(\ref{eq:lmoc}), we still require a way to estimate properly
the conditional probabilities $u^{[\alpha]}, u^{[\beta]},
v^{[\alpha]}, v^{[\beta]},  u^{[\alpha]}$, and $z^{[\beta]}$.
Following a similar approach as the one described above, we have that
\begin{equation}
\begin{array}{ll}
u^{[\alpha]}=& \big[1-G_1^{[\alpha]}(1-\phi
u^{[\alpha]}-(1-\phi)v^{[\alpha]}) \big] \times
\\
&	   \big[1-G_0^{[\beta]}(1-\phi
           u^{[\beta]}-(1-\phi)v^{[\beta]}) \big]
\end{array} \; ,
\label{eq:u}
\end{equation}
\begin{equation}
\begin{array}{ll}
v^{[\alpha]}=& \big[1-G_1^{[\alpha]}(1-\phi
u^{[\alpha]}-(1-\phi)z^{[\alpha]}) \big] \times
\\
&	   \big[1-G_0^{[\beta]}(1-\phi
  u^{[\beta]}-(1-\phi)z^{[\beta]}) + 
\\
& - G_0^{[\beta]}(1-\phi) +G_0^{[\beta]}((1-\phi)(1-z^{[\beta]}))  \big]
\end{array} \; ,
\label{eq:v}
\end{equation}
and
\begin{equation}
\begin{array}{ll}
z^{[\alpha]}=&  \big[1-G_1^{[\alpha]}(1-\phi
  u^{[\alpha]}-(1-\phi)z^{[\alpha]}) + 
\\
& - G_1^{[\alpha]}(1-\phi) +G_1^{[\alpha]}((1-\phi)(1-z^{[\alpha]}))
  \big] \times
\\
&	   \big[1-G_0^{[\beta]}(1-\phi
  u^{[\beta]}-(1-\phi)z^{[\beta]}) + 
\\
& - G_0^{[\beta]}(1-\phi) +G_0^{[\beta]}((1-\phi)(1-z^{[\beta]}))  \big]
\end{array} \; .
\label{eq:z}
\end{equation}
The intuition behind the previous equations is
straightforward. When looking at an edge on layer $\alpha$, the
probability
that this edge will bring to the LMOC will depend on the generating
function of the excess degree
distribution of the layer, i.e., $G_1^{[\alpha]}(x) =
[d/dx G_0^{[\alpha]}(x)] / [d/dx G_0^{[\alpha]}(1)]$. On the other
hand, the probability that a node at the end of this edge is also attached 
to the LMOC on layer
$\beta$
will depend only on the degree of the node itself on layer $\beta$, accounted
by the generating function of the degree
distribution, namely  $G_0^{[\beta]}(x)$.
Note that equations for
$u^{[\beta]}, v^{[\beta]}$, and $z^{[\beta]}$ can be obtained by simply swapping
$\alpha$ and $\beta$ in Eqs.~(\ref{eq:u}), ~(\ref{eq:v}),
and ~(\ref{eq:z}).

At this point, we have everything necessary to
estimate Eq.~(\ref{eq:lmoc}). We have first to solve 
Eqs.~(\ref{eq:u}), ~(\ref{eq:v}),
and ~(\ref{eq:z}), and the analogous ones for layer $\beta$, by iteration.
Then, we can plug the obtained values of the conditional probabilities
for generic edges into the equations for the nodes.

\begin{figure}[!htb]
\begin{center}
\includegraphics[width=0.4\textwidth]{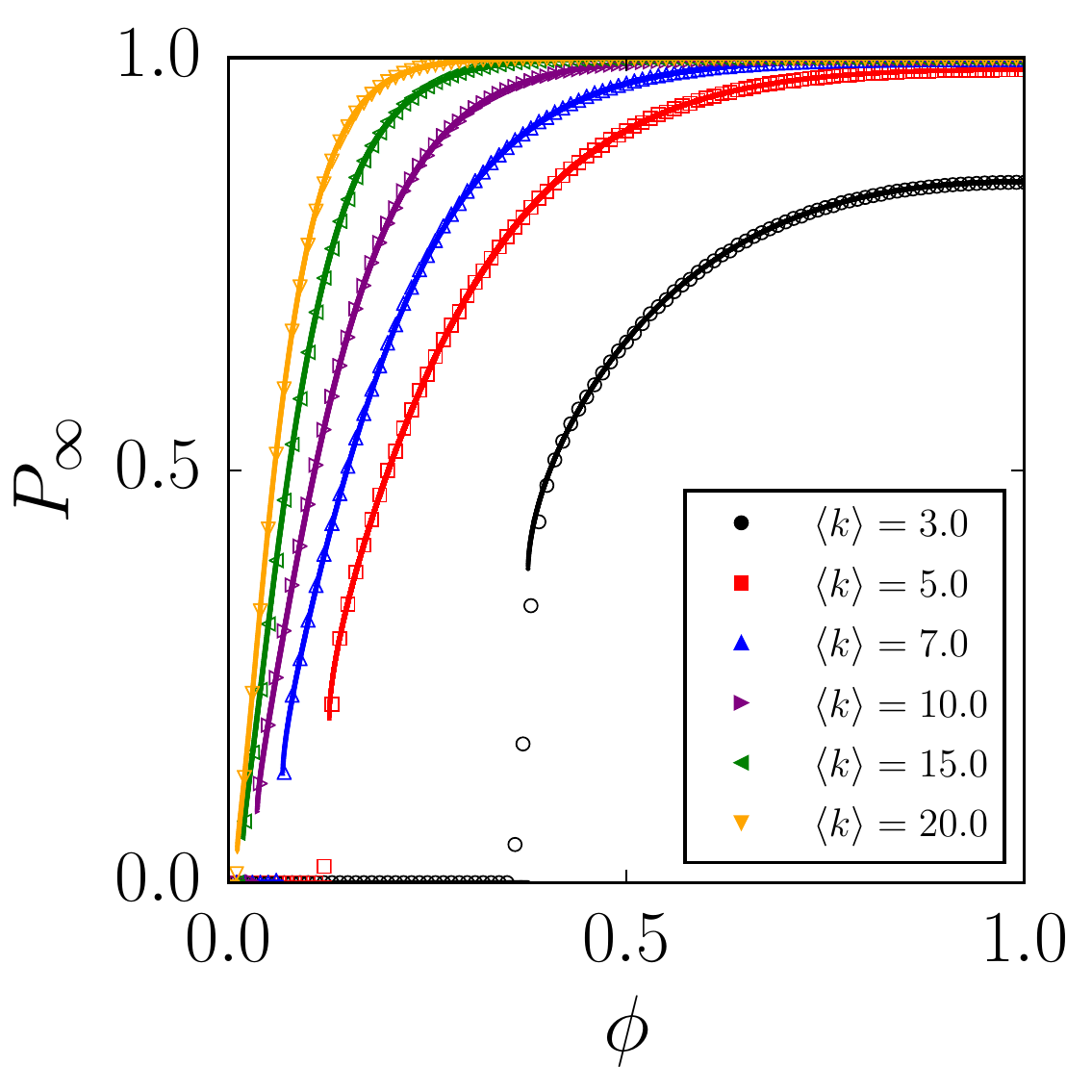}
\end{center}
\caption{Observability transition in random Poisson multiplex networks. 
We consider multiplex networks with layers generated independently
and given by realizations of the configuration model with $N=10, 000$ nodes and
degree sequence generated according to a Poisson distribution
with average degree $\langle k \rangle$. Results of our theoretical
method (small symbols) are compared with those of numerical
simulations of the observability model (large symbols). We consider
different
values of $\langle k \rangle$, $\langle k \rangle = 3.0, 5.0, 7.0, 10.0,
15.0$, and $20.0$. Values of $P_\infty$ for the various cases
drop to zero in the same order if the figure is read from right to left.}
\label{fig:1}
\end{figure}

\begin{figure}[!htb]
\begin{center}
\includegraphics[width=0.4\textwidth]{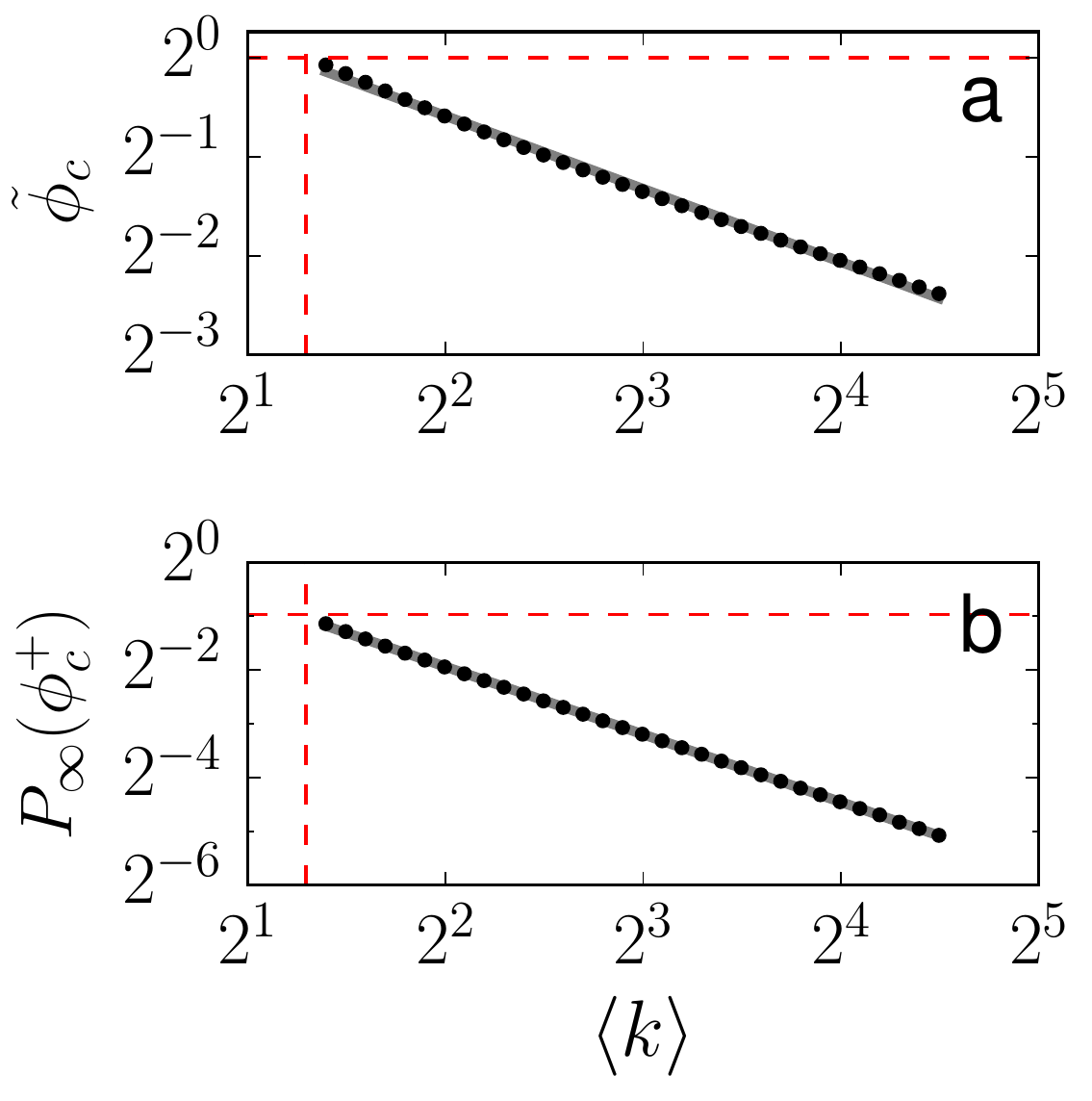}
\end{center}
\caption{Observability transition in random Poisson multiplex networks. 
(a) We plot the rescaled value of the critical threshold 
$\tilde{\phi}_c = \phi_c + (1 - \phi_c) [1 - (1 - \phi_c)^{\langle k 
  \rangle }]$ (black circles) as a function of the average degree 
$\langle k \rangle$ for Poisson multiplex networks of infinite size. 
We find that $\tilde{\phi}_c \propto \langle k \rangle^{-3/4}$ (gray 
full line). 
The vertical red line indicates the value $\langle k \rangle_c \simeq 
2.4554$
where $\tilde{\phi}_c = 1$ (horizontal red line).  
(b) Height of the discontinuous jump $P_\infty(\phi_c^+)$ 
as a function of the average degree 
$\langle k \rangle$ (black circles) for the same multiplex networks of panel (a). 
We find that $P_\infty(\phi_c^+) \propto \langle k \rangle^{-7/4}$
(gray full line). The vertical red line indicates the value $\langle k \rangle_c \simeq 
2.455$
where $P_\infty(\phi_c^+) \simeq 0.511$ (horizontal red line).}
\label{fig:1a}
\end{figure}

\begin{figure}[!htb]
\begin{center}
\includegraphics[width=0.4\textwidth]{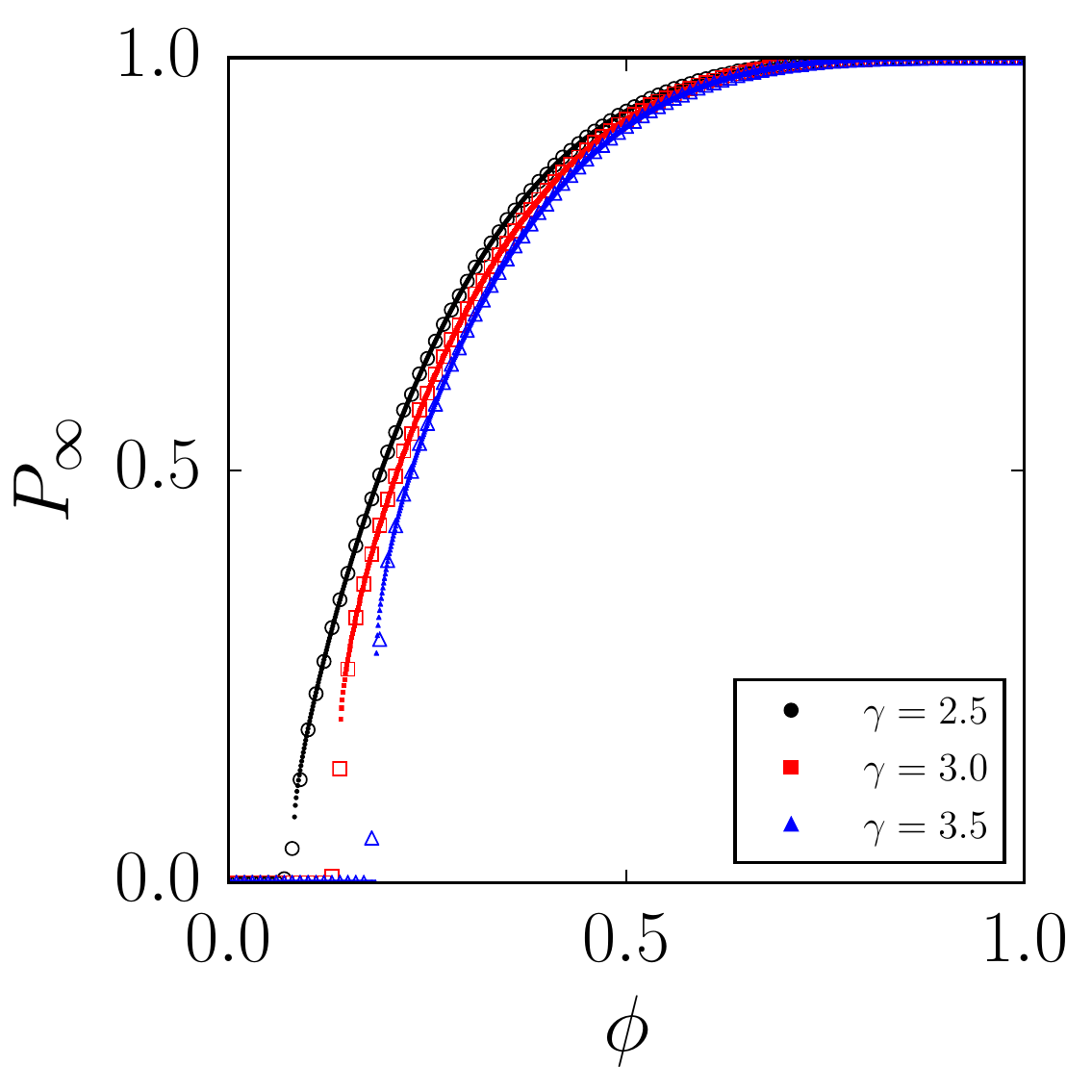}
\end{center}
\caption{Observability transition in a single random scale-free multiplex networks. 
We consider multiplex networks with layers generated independently 
and given by realizations of the configuration model with $N=10, 000$ nodes and 
degree sequence generated according to a power-law degree distribution 
$P^{[\alpha]}(k) = P^{[\beta]}(k) \sim k^{-\gamma}$
with degree exponent $\gamma$ (minimal degree is set $k_{min}=3$, so that
edge overlap between layers is negligible).
Results of message passing framework
without overlap (small symbols) are compared with those of numerical 
simulations of the observability model (large symbols). We consider 
different 
values of $\gamma$, $\gamma = 2.5, 3.0$ and $3.5$. 
Values of $P_\infty$ for the various cases 
drop to zero in the same order if the figure is read from right to left.}
\label{fig:3}
\end{figure}
Figure~\ref{fig:1} shows a comparison between the results
of direct numerical simulations of the observability model
and the numerical solutions of our equations for duplex networks
formed by layers obeying  Poisson degree distributions, i.e.,
$P^{[\alpha]}(k) = P^{[\beta]}(k) = \frac{\langle k \rangle^{k} e^{-
    \langle k \rangle}}{\langle k \rangle !}$. The figure provides
evidence of a perfect agreement between theory and simulations.
As the generating functions $G_0^{[\alpha]}$, $G_1^{[\alpha]}$,
$G_0^{[\beta]}$, and $G_1^{[\beta]}$ can be written in a closed form
for the Poisson distribution, we can also study the numerical solution
of the equations for infinitely large networks (Fig.~\ref{fig:1a}).
No transition takes place for $\langle k \rangle < 2.4554$, i.e.,
the
same value found for the standard percolation
model~\cite{buldyrev2010catastrophic}.  This is
expected from the equivalence between between the observability model
and
the standard percolation model for $\phi = 1$.
For $\langle k \rangle \geq 2.4554$, the behavior of the observability
model
becomes different from the one observed in site
percolation.
There, we have that the critical occupation probability $p_c$
and the height of the discontinuous jump $P_\infty (p_c^{+})$ decay
to zero as a function of the average degree $\langle k
\rangle$ as  $p_c \propto P_\infty (p_c^{+}) \propto \langle k
\rangle^{-1}$~\cite{buldyrev2010catastrophic}.
For the observability model, 
we find instead that the rescaled
critical threshold $\tilde{\phi}_c = \phi_c + (1 - \phi_c) [1 - (1 - \phi_c)^{\langle k
  \rangle}]$ doesn't decrease as $\langle k
\rangle^{-1}$ as a naive 
mapping between the two models would predict. Instead, the scaling is
compatible with $\tilde{\phi}_c \propto \langle k \rangle^{-3/4}$
(Fig.~\ref{fig:1a}a). The height of the discontinuous jump
$P_\infty(\phi_c^+)$ also doesn't go to zero as $\langle k
\rangle^{-1}$
as predicted in the standard percolation model, but instead
as $P_\infty(\phi_c^+) \propto \langle k \rangle^{-7/4}$
(Fig.~\ref{fig:1a}b).

\section{Observability transition in real multiplex networks}
\label{sec:real}
In this section, we develop an analytic framework able to
approximate the phase diagram of the observability transition for
a duplex with given
adjacency matrices for the two layers. This information is encoded in
the sets $\partial_i^{[\alpha]}$ and $\partial_i^{[\beta]}$ containing
the neighbors of  node $i$ respectively in layers $\alpha$ and $\beta$
for every node $i$. We indicate the sizes of these sets
respectively as $k^{[\alpha]}_i$ and $k^{[\beta]}_i$.  We make use of two assumptions: (i) layers have
null overlap in the sense that set of  neighbors of the same node in
the two layers have null intersection; (ii) neighbors of a
node are not attached to each other.

Suppose we are interested in estimating the
probability $s_i$ that node $i$ is part of the LMOC. This will happen
if node $i$ is receiving at least 
one message in layer $\alpha$ and one message in layer $\beta$
about the belonging to the LMOC. We consider average message values
over the ensemble of random placements
of observers in the network. In particular, we define
three different messages for every edge $j \to i$ in layer
$\alpha$ dependent on the states of the nodes $j$ and $i$:

\begin{enumerate}

\item $u^{[\alpha]}_{j\to i}$, that is the average value of the message that node $j$ sends
  to node $i$, when $j$ is directly observable.

\item $v^{[\alpha]}_{j\to i} $, that is the average value of the message that node $j$ sends
  to node $i$, when $j$ is not directly observable.

\item $z^{[\alpha]}_{j\to i}$, that is the average value of the message that node $j$ sends
  to node $i$, when neither $j$ nor $i$ are directly observable.

\end{enumerate}

$u^{[\beta]}_{j\to i}$, $v^{[\beta]}_{j\to i}$, and $z^{[\beta]}_{j\to
  i}$ 
represent the same quantities as above but for layer $\beta$.
Note that messages traveling on the same edge
but in opposite direction are different.

The generic node $i$ is part of the LMOC if one of these two conditions are
met: (i) the node is directly observable and attached to at least one
other node in layer $\alpha$ and layer $\beta$ that are part of the LMOC; (ii) the node is indirectly
observable and attached to at least one
other node in layer $\alpha$ and layer $\beta$ that are part of the LMOC.

Considering only layer $\alpha$, for case (i), we can write

\begin {equation}
\begin{array}{ll}
	q^{[\alpha]}_i= & \bigg[1-
	\prod\limits_{j \in \partial_i^{[\alpha]}} (1-\phi
  u^{[\alpha]}_{j\to i}-(1-\phi)v^{[\alpha]}_{j \to i})
	\bigg]
\end{array} \; .
\label{eq:q_a}
\end {equation}

For case (ii), we have instead
\begin {equation}
\begin{array}{ll}
	r^{[\alpha]}_i= & \bigg[1-
	\prod\limits_{j \in \partial_i^{[\alpha]}} (1-\phi
  u^{[\alpha]}_{j\to i}-(1-\phi)z^{[\alpha]}_{j \to i}) +
\\ 
&
- (1-\phi)^{k_i^{[\alpha]}} 
\big[1-\prod\limits_{j \in \partial_i^{[\alpha]}} (1-z^{[\alpha]}_{j\to
  i}) \big] \bigg] \times
\end{array} \; .
\label{eq:r_a}
\end {equation}
The same type of equations are valid for layer $\beta$. As the node
must be part of the LMOC in both layer simultaneously, we can
write 
\begin {equation}
	q_i= \phi \, q^{[\alpha]}_i q^{[\beta]}_i
\; ,
\label{eq:q_b}
\end {equation}
and
\begin {equation}
	r_i= (1- \phi) \, r^{[\alpha]}_i r^{[\beta]}_i
\; ,
\label{eq:r_b}
\end {equation}
where the factors $\phi$ and $1-\phi$ come from the fact that we are
considering the cases of a directly observable and a non directly
observable node, respectively. The size of the average
LMOC is given by
\begin{equation}
P_\infty =  \frac{1}{N} \, \sum_i q_i + r_i \; .
\label{eq:lmoc_a}
\end{equation}
The self-consistent equations for the messages are instead
given by
\begin {equation}
\begin{array}{l}
u^{[\alpha]}_{j \to i} = \bigg[1-
	\prod_{k \in \partial_j^{[\alpha]} \setminus i} (1-\phi
        u^{[\alpha]}_{k\to j}-(1-\phi)v^{[\alpha]}_{k\to j})
	\bigg] \times \\
	\bigg[1-
	\prod_{k \in \partial_j^{[\beta]} }(1-\phi u^{[\beta]}_{k\to
            j}-(1-\phi)v^{[\beta]}_{k\to j})
	\bigg]
\end{array} \; ,
\label{eq:u_a}
\end {equation}

\begin{equation}
\begin{array}{l}
v^{[\alpha]}_{j \to i} = 
\bigg[1-
	\prod_{k \in \partial_j^{[\alpha]} \setminus i} (1-\phi
  u^{[\alpha]}_{k\to j}-(1-\phi)z^{[\alpha]}_{k\to j})
	\bigg] \times \\
	\bigg[
	1-\prod\limits_{k \in \partial_j^{[\beta]}} (1-\phi
  u^{[\beta]}_{k\to j}-(1-\phi)z^{[\beta]}_{k\to j}) +
\\
-(1-\phi)^{k_j^{[\beta]}} 
	\big[1-\prod\limits_{k \in \partial_j^{[\beta]}} (1-z^{[\beta]}_{k\to j}) \big]
	\bigg]
\end{array} \; ,
\label{eq:v_a}
\end {equation}
and
\begin{equation}
\begin{array}{l}

z^{[\alpha]}_{j \to i} = 
\bigg[
	1-\prod\limits_{k \in \partial_j^{[\alpha]} \setminus i}
  (1-\phi u^{[\alpha]}_{k\to j}-(1-\phi)z^{[\alpha]}_{k\to j})
+
\\
-(1-\phi)^{k_j^{[\alpha]}-1} 
	\big[1-\prod\limits_{k \in \partial_j^{[\alpha]} \setminus i}
  (1-z^{[\alpha]}_{k\to j}) \big]
	\bigg]  \times \\

	\bigg[
	1-\prod\limits_{k \in \partial_j^{[\beta]}}
  (1-\phi u^{[\beta]}_{k\to j}-(1-\phi)z^{[\beta]}_{k\to j})
+
\\
-(1-\phi)^{k_j^{[\beta]}} 
	\big[1-\prod\limits_{k \in \partial_j^{[\beta]} }
  (1-z^{[\beta]}_{k\to j}) \big]
	\bigg]
\end{array} \; ,
\label{eq:z_a}
\end {equation}
for a generic edge $j \to i$ belonging to layer $\alpha$.
Equations with the same structure allow to compute the
probabilities defined for edges in layer $\beta$. We remark that the
equations above make use of the locally treelike approximation,
hence the products appearing on their right-hand sides.
Further, backtracking terms are excluded in the products.

The analytic framework is now completed. To estimate
the average size of the LMOC as a function of $\phi$, one needs to:
first, solve Eqs.~(\ref{eq:u_a}), ~(\ref{eq:v_a}), and ~(\ref{eq:z_a})
by iteration; second, plug these solutions into Eqs.~(\ref{eq:q_a}),
and ~(\ref{eq:r_a}); third, estimate in sequence Eqs.~(\ref{eq:q_b}),
~(\ref{eq:r_b}), and~(\ref{eq:lmoc_a}).

We performed comparisons between numerical solutions of the framework
and results of numerical simulations for a single multiplex networks formed by
random scale-free network layers which have negligible overlap (see Fig.~\ref{fig:3}).
The agreement between the two approaches is remarkable.

We stress the fact that the framework presented above
is valid under the assumption that the two layers that compose the
multiplex do not share any edge. This is a very strong assumption,
often violated by real-world networks. 
For instance, edge overlap
may have 
a dramatic consequence on the properties of the percolation transition~\cite{bianconi2013statistical,
  baxter2016correlated, min2015link, radicchi2015percolation, 
PhysRevE.94.032301, PhysRevE.94.060301}.
We developed a mathematical framework valid 
in case of edge overlap. Given its length and complexity,
we present the details only in the Supplemental Material.
\begin{figure*}[!htb]
\begin{center}
\includegraphics[width=0.95\textwidth]{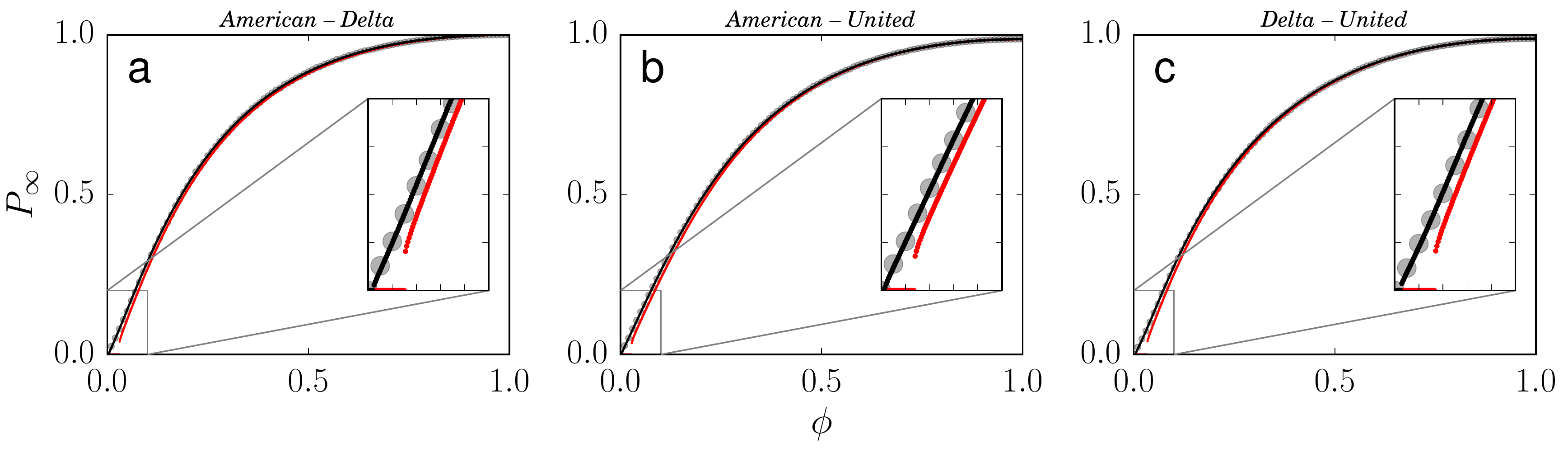}
\end{center}
\caption{Observability transition in the US air transportation
  multiplex network. 
(a) The system is obtained by combining American Airlines and Delta
routes. We consider only US domestic flights operated in January,
2014, and construct the duplex network where airports are nodes and
connections on the layers are determined by the existence of at least
a flight between the two locations. In the diagram, the
gray big circles represents results of numerical simulations, the red
small circles stand for results from the framework that doesn't
account
for edge overlap, and
the black small circles represent results obtained from the
mathematical
framework that accounts for edge overlap. The inset shows a zoom of
a specific part of the diagram.
(b) Same as in panel (a), but for the combination of American
Airlines and United flights. (c) Same as in panel (a), but for the
combination of Delta and United flights.}
\label{fig:air}
\end{figure*}
\begin{figure*}[!htb]
\begin{center}
\includegraphics[width=0.95\textwidth]{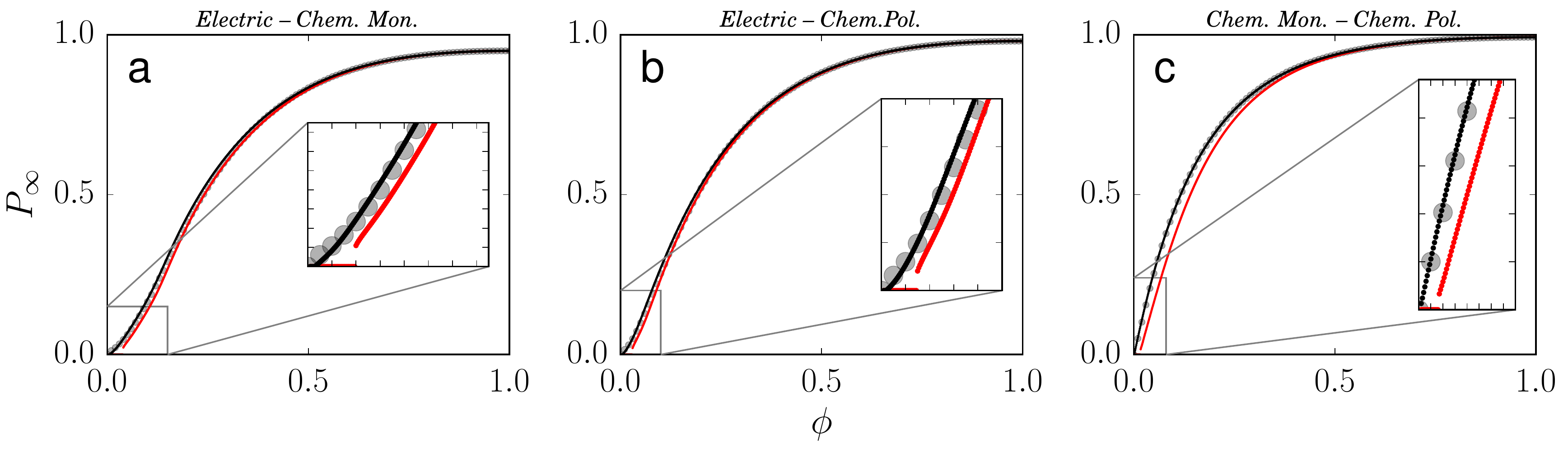}
\end{center}
\caption{Observability transition in the {\it
    C. Elegans} connectome
  multiplex network. Edges in different layers represent different 
types of synaptic junctions among the neurons: electrical, chemical
monadic, 
and chemical polyadic.
(a) Analysis of the multiplex obtained by combining together the
layers of electrical and chemical monadic interactions. In the diagram, the
gray big circles represents results of numerical simulations, the red
small circles stand for results from the framework that doesn't
account
for edge overlap, and
the black small circles represent results obtained from the
mathematical
framework that accounts for edge overlap. The inset shows a zoom of
a specific part of the diagram.
(b) Same as in panel (a), but for the combination of
electrical and chemical polyadic interactions. (c) Same as in panel (a), but for the
combination of chemical monadic and polydiatic interactions.}
\label{fig:cele}
\end{figure*}
Indeed, results from the application of the two methods (with or without edge
overlap)
to real-world multiplex networks provide very different scenarios and
degree of accuracy. In Figure~\ref{fig:air} for example, we show a
comparisons between numerical simulations and numerical solutions of
the
frameworks when applied to a multiplex transportation network~\cite{radicchi2015percolation}.
The framework that accounts for edge overlap well approximates
 the ground-truth results obtained with numerical simulations.
The observability transitions appear smooth, and the transition point
is very close to zero. The framework that doesn't account for edge
overlap
instead shows an abrupt transition, and a threshold larger than zero.
Similar considerations are valid also for the multiplex network
representing the {\it
    C. Elegans} connectome (Fig.~\ref{fig:cele})~\cite{chen2006wiring,
    de2014muxviz}.

\section{Conclusions}
\label{sec:conclusions}
In this paper, we extended 
the observability model, originally considered
on isolated networks, to multiplex networks.
In particular, we focused our attention on the emergence 
of the largest cluster of mutually observed nodes
as a function of the microscopic probability
of individual nodes to host observers. 
We developed  mathematical frameworks able
to well describe the observability diagram in synthetic and real-world
multiplex networks. Our results indicate that
the features of this phase transition cannot be 
trivially deduced from those
valid for the percolation transition.
This statement is true both for randomly generated multiplex 
networks, as well as for multiplex networks 
representing real-world systems.
Interestingly, real-world multiplex networks seem
to be always in the observable regime, as long as the
fraction of nodes that is directly observed is larger than zero.
This fact seems due to the natural, and ubiquitously observed,
presence of edge overlap among the layers that compose real
multiplex networks.

\begin{acknowledgments}
FR acknowledges 
support from the National Science 
Foundation (Grant CMMI-1552487) and 
the US Army Research Office (W911NF-16-1-0104).
\end{acknowledgments}



\clearpage

\newpage

 \setcounter{page}{1}
\renewcommand{\theequation}{SM\arabic{equation}}
\setcounter{equation}{0}
\renewcommand{\thefigure}{SM\arabic{figure}}
\setcounter{figure}{0}
\renewcommand{\thetable}{SM\arabic{table}}
\setcounter{table}{0}

\onecolumngrid

\section*{Supplemental Material}

\subsection*{Message passing with edge overlap}
\label{sec:overlap}
This section is devoted to the description of the message-passing
algorithm to approximate the behaviour of the Largest 
Mutually Observable Cluster (LMOC) in multiplex
networks. The formalism differ from the one already presented in the
main text for the fact that the only approximation used is the locally
treelike ansatz. Edge overalp among layers is instead accounted by
this framework. The method presented here is
a generalization of the algorithm proposed in
Ref.~\cite{PhysRevE.94.060301}
for standard site percolation on multiplex networks.
The method relies on the definition of the multilink $\vec{m}_{ji} =
(a^{[\alpha]}_{ji}, a^{[\beta]}_{ji})$ for
every pairs of nodes $j$ and $i$ in the duplex, where
 $a^{[\alpha]}_{ji}=1$ if the nodes are connected in layer $\alpha$,
 and $a^{[\alpha]}_{ji}=0$, otherwise. The same definition applies for 
layer $\beta$. Messages considered in the approach are:

\begin{enumerate}

\item $u_{j \to i}^{\vec{m}_{ji},\vec n}$, valid when node~$j$ is directly observable.

\item $v_{j \to i}^{\vec{m}_{ji},\vec n}$, valid when node~$j$ is not directly observable.

\item $z_{j \to i}^{\vec{m}_{ji},\vec n}$, valid  when neither node~$i$ nor node~$j$ are directly observable.

\end{enumerate}
In the definition of the messages, we use
$\vec n = (n^{[\alpha]}, n^{[\beta]})$ to maintain the notation as
compact as possible. Different message are indicated by different
values of $\vec n$, namely $(0, 0)$, $(1,0)$, $(0,
1)$, and $(1,1)$. Note that if $n^{[\alpha]} (1 - a_{ji}^{[\alpha]}) +
n^{[\beta]} (1 - a_{ji}^{[\beta]})  \neq 0$, the corresponding message
is automatically zero. Further, normalization implies that $u_{j \to
  i}^{\vec{m}_{ji},(0, 0)} = 1 - u_{j \to
  i}^{\vec{m}_{ji},(1, 0)} - u_{j \to
  i}^{\vec{m}_{ji},(0, 1)} - u_{j \to
  i}^{\vec{m}_{ji},(1, 1)}$. 
It is further convenient to cumulate messages over layers as:
\[
u_{j \to{} i}^{[\alpha]} = u_{j \to{} i}^{\vec{m}_{ji},(1,0)} + 
					   u_{j \to{}
                                             i}^{\vec{m}_{ji},(1,1)}
                                           \; ,
\]
\[
u_{j \to{} i}^{[\beta]} = u_{j \to{} i}^{\vec{m}_{ji},(0,1)} + 
					   u_{j \to{}
                                             i}^{\vec{m}_{ji},(1,1)}
                                           \; ,
\]
and
\[
					   u_{j \to{} i}^{[\alpha, \beta]} = u_{j \to{} i}^{\vec{m}_{ji},(1,0)} + 
						 u_{j \to{} i}^{\vec{m}_{ji},(0,1)} +
					     u_{j \to{}
                                               i}^{\vec{m}_{ji},(1,1)}
                                             \; .
\]

The same definitions are  valid for $u$- and $z$-type messages.

The probability that node~$i$ is in the LMOC is calculated
in different manner depending on whether node $i$ is
(i) directly observable or (ii) not directly observable.
For case (i), we have

\begin{align}
q_i =  
\phi \Bigg[1-\bigg[ \prod_{j\in\partial_i} \big(1-\phi u_{j\to{}i}^{[\alpha]}-(1-\phi) v_{j\to{}i}^{[\alpha]} \big) \bigg] 
 - \bigg[ \prod_{j\in\partial_i} \big(1-\phi u_{j\to{}i}^{[\beta]}-(1-\phi) v_{j\to{}i}^{[\beta]} \big) \bigg] +
 \bigg[ \prod_{j\in\partial_i} \big(1-\phi u_{j\to{}i}^{[\alpha,
  \beta]}-(1-\phi) v_{j\to{}i}^{[\alpha, \beta]} \big) \bigg] \Bigg] \;.
\end{align} 
Essentially, the probability for node $i$ to be directly observable
and
part of the LMOC is given by the product of the probability
to be directly observable and attached at least to another
node that is part of the LMOC. The latter is estimated as
one minus the probability that none of the nodes connected to $i$
are part of the LMOC.

To account for the overlap among edges 
in the two layers, we need to make a distinction among neighbors of a node.
We define three sets of neighbors for node~$i$:
$\partial_i^{[\alpha]}$, that is set of nodes that
are neighbors of node~$i$ just in layer~$\alpha$;
$\partial_i^{[\beta]}$, that is set of nodes that
are neighbors of node~$i$ just in layer~$\beta$;
$\partial_i^{[\alpha,\beta]}$, that is the set of nodes that are
neighbors of node~$i$ at the same time in both layers. 
In particular, we indicate with $k_i^{[\alpha]}$, $k_i^{[\beta]}$,
$k_i^{[\alpha, \beta]}$ the size of 
the sets $\partial_i^{[\alpha]}$, $\partial_i^{[\beta]}$,
$\partial_i^{[\alpha,\beta]}$, respectively.  We also define degree
of each node in layer $\alpha$, layer $\beta$ and total as follows:
$k_i^{[1]}=k_i^{[\alpha]}+k_i^{[\alpha, \beta]}$, 
$k_i^{[2]}=k_i^{[\beta]}+k_i^{[\alpha, \beta]}$, 
$k_i^{[1,2]}=k_i^{[\alpha]}+k_i^{[\beta]}+k_i^{[\alpha, \beta]}$.


\begin{align}
r_i = & (1-\phi) \big[ 1 - A - B + C \big]
\end{align}

\begin{align}
 A = &\bigg[ \prod_{j\in\partial_i} \big(1-\phi u_{j\to{}i}^{[\alpha]}-(1-\phi) z_{j\to{}i}^{[\alpha]} \big) \bigg]+
(1-\phi)^{k_i^{[1]}} \bigg[ 1-\prod_{j\in\partial_i}(1-z_{j\to{}i}^{[\alpha]}) \bigg] \nonumber\\
 B = &\bigg[ \prod_{j\in\partial_i} \big(1-\phi u_{j\to{}i}^{[\beta]}-(1-\phi) z_{j\to{}i}^{[\beta]} \big) \bigg]+
(1-\phi)^{k_i^{[2]}} \bigg[
       1-\prod_{j\in\partial_i}(1-z_{j\to{}i}^{[\beta]}) \bigg]
       \nonumber\\
 C = &\bigg[ \prod_{j\in\partial_i} \big(1-\phi u_{j\to{}i}^{[\alpha, \beta]}-(1-\phi) z_{j\to{}i}^{[\alpha, \beta]} \big) \bigg]+
(1-\phi)^{k_i^{[1, 2]}} \bigg[ 1-\prod_{j\in\partial_i}(1-z_{j\to{}i}^{[\alpha, \beta]}) \bigg] +\nonumber\\
& \bigg[ \prod_{j\in\partial_i^{[\alpha]}} \big(1-\phi u^{[\alpha]}_{j\to{}i}-(1-\phi) z^{[\alpha]}_{j\to{}i} \big) -
(1-\phi)^{k_i^{[\alpha]}}  \prod_{j\in\partial_i^{[\alpha]}}(1-z^{[\alpha]}_{j\to{}i}) \bigg] \times \nonumber\\
& (1-\phi)^{k_i^{[2]}} \bigg[ \prod_{j\in\partial_i^{[\alpha,\beta]}}(1-z_{j\to{}i}^{[\alpha]}) -
\prod_{j\in\partial_i^{[\beta]}}(1-z^{[\beta]}_{j\to{}i}) 
\prod_{j\in\partial_i^{[\alpha,\beta]}}(1-z_{j\to{}i}^{[\alpha, \beta]})\bigg] +
\nonumber\\
& \bigg[ \prod_{j\in\partial_i^{[\beta]}} \big(1-\phi u^{[\beta]}_{j\to{}i}-(1-\phi) z_{j\to{}i}^{[\beta]} \big) -
(1-\phi)^{k_i^{[\beta]}}  \prod_{j\in\partial_i^\beta}(1-z^{[\beta]}_{j\to{}i}) \bigg] \times \nonumber\\
& (1-\phi)^{k_i^{[1]}} \bigg[ \prod_{j\in\partial_i^{[\alpha,\beta]}}(1-z_{j\to{}i}^{[\beta]}) -
\prod_{j\in\partial_i^{[\alpha]}}(1-z^{[\alpha]}_{j\to{}i}) 
\prod_{j\in\partial_i^{[\alpha,\beta]}}(1-z_{j\to{}i}^{[\alpha, \beta]})\bigg]
\end{align} 

The terms$A$ and $B$ are obvious. $C$ instead contains a series of
exceptions
that must be handled to properly account for overlap. In particular,
\begin{enumerate}
\item The first term in $C$ stands for the probability that node $i$
  is not connected to the LMOC in none of the layers. It accounts also
  for the exception where all of its neighbors are not directly observable.

\item Probability for node $i$ of being not connected to LMOC through
  layer $\alpha$ and  
being not observable in layer $\beta$. Connections in layer $\beta$ do
not matter, but overlapping links, may connect node $i$ to the LMOC in
layer $\alpha$.

\item Same as point 2, but swapping $\alpha$ with $\beta$.

\end{enumerate}

The exceptions mentioned above are computed noting that

\begin{align}
 \prod_{j\in\partial_i} \big(1-\phi u_{j\to{}i}^{[\alpha, \beta]}-(1-\phi) v_{j\to{}i}^{[\alpha, \beta]} \big) = 
 & \bigg[ \prod_{j\in\partial_i^{[\alpha]}} \big(1-\phi u_{j\to{}i}^{(1,0),(1,0)}-(1-\phi) v_{j\to{}i}^{(1,0),(1,0)} \big) \bigg] \times \nonumber\\
 & \bigg[ \prod_{j\in\partial_i^{[\beta]}} \big(1-\phi u_{j\to{}i}^{(0,1),(0,1)}-(1-\phi) v_{j\to{}i}^{(0,1),(0,1)} \big) \bigg] \times \nonumber\\
  &\bigg[ \prod_{j\in\partial_i^{[\alpha,\beta]}} \big(1-\phi
    u_{j\to{}i}^{[\alpha, \beta]}-(1-\phi) v_{j\to{}i}^{[\alpha,
    \beta]} \big) \bigg] \; .
\end{align}

We can now derive self-consistent equations for the messages.
For $u$- and $v$-type messages, these are provided by the following equations:
\begin{align}
u_{j\to{}i}^{(1,1),(1,1)}=& u_{j\to{}i}^{(1,0),(1,0)}=u_{j\to{}i}^{(0,1),(0,1)} = 1-\bigg[ \prod_{k\in\partial_j \setminus i} \big(1-\phi u_{k\to{}j}^{[\alpha]}-(1-\phi) v_{k\to{}j}^{[\alpha]} \big) \bigg] \nonumber\\
& - \bigg[ \prod_{k\in\partial_j \setminus i} \big(1-\phi u_{k\to{}j}^{[\beta]}-(1-\phi) v_{k\to{}j}^{[\beta]} \big) \bigg] +
 \bigg[ \prod_{k\in\partial_j \setminus i} \big(1-\phi
  u_{k\to{}j}^{[\alpha, \beta]}-(1-\phi) v_{k\to{}j}^{[\alpha, \beta]}
  \big) \bigg] \; ,
\end{align} 

\begin{align}
u_{j\to{}i}^{(1,1),(1,0)} =& \bigg[ \prod_{k\in\partial_j \setminus i} \big(1-\phi u_{k\to{}j}^{[\beta]}-(1-\phi) v_{k\to{}j}^{[\beta]} \big) \bigg] -
 \bigg[ \prod_{k\in\partial_j \setminus i} \big(1-\phi
                             u_{k\to{}j}^{[\alpha, \beta]}-(1-\phi)
                             v_{k\to{}j}^{[\alpha, \beta]} \big)
                             \bigg] \; ,
\end{align} 

\begin{align}
u_{j\to{}i}^{(1,1),(0,1)} = &\bigg[ \prod_{k\in\partial_j \setminus i} \big(1-\phi u_{k\to{}j}^{[\alpha]}-(1-\phi) v_{k\to{}j}^{[\alpha]} \big) \bigg]- 
 \bigg[ \prod_{k\in\partial_j \setminus i} \big(1-\phi
                              u_{k\to{}j}^{[\alpha, \beta]}-(1-\phi)
                              v_{k\to{}j}^{[\alpha, \beta]} \big)
                              \bigg] \; ,
\end{align}

\begin{align}
v_{j\to{}i}^{(1,1),(1,1)}= 1-& \bigg[ \prod_{k\in\partial_j \setminus i} \big(1-\phi u_{k\to{}j}^{[\alpha]}-(1-\phi) z_{k\to{}j}^{[\alpha]} \big) \bigg]- \bigg[ \prod_{k\in\partial_j \setminus i} \big(1-\phi u_{k\to{}j}^{[\beta]}-(1-\phi) z_{k\to{}j}^{[\beta]} \big) \bigg] +
\nonumber \\
& \bigg[ \prod_{k\in\partial_j \setminus i} \big(1-\phi
  u_{k\to{}j}^{[\alpha, \beta]}-(1-\phi) z_{k\to{}j}^{[\alpha, \beta]}
  \big) \bigg] \; ,
\end{align}

\begin{align}
v_{j\to{}i}^{(1,1),(1,0)} =& \bigg[ \prod_{k\in\partial_j \setminus i} \big(1-\phi u_{k\to{}j}^{[\beta]}-(1-\phi) z_{k\to{}j}^{[\beta]} \big) \bigg] -
 \bigg[ \prod_{k\in\partial_j \setminus i} \big(1-\phi
                             u_{k\to{}j}^{[\alpha, \beta]}-(1-\phi)
                             z_{k\to{}j}^{[\alpha, \beta]} \big)
                             \bigg] \; ,
\end{align} 

\begin{align}
v_{j\to{}i}^{(1,1),(0,1)} = &\bigg[ \prod_{k\in\partial_j \setminus i} \big(1-\phi u_{k\to{}j}^{[\alpha]}-(1-\phi) z_{k\to{}j}^{[\alpha]} \big) \bigg]- 
 \bigg[ \prod_{k\in\partial_j \setminus i} \big(1-\phi
                              u_{k\to{}j}^{[\alpha, \beta]}-(1-\phi)
                              z_{k\to{}j}^{[\alpha, \beta]} \big)
                              \bigg] \; ,
\end{align} 

and

\begin{align}
v_{j\to{}i}^{(1,0),(1,0)}= 1-& \Bigg[ \prod_{k\in\partial_j \setminus i} \big(1-\phi u_{k\to{}j}^{[\alpha]}-(1-\phi) z_{k\to{}j}^{[\alpha]} \big) \Bigg] -  \Bigg[ \prod_{k\in\partial_j \setminus i} \big(1-\phi u_{k\to{}j}^{[\beta]}-(1-\phi) z_{k\to{}j}^{[\beta]} \big) + (1-\phi)^{k_j^{[2]}} \times
\nonumber\\ 
& \bigg[1-\prod_{k\in\partial_j \setminus i} (1-z_{k\to{}j}^{[\beta]}) \bigg] \Bigg] +
\nonumber \\
& \Bigg[ \prod_{k\in\partial_j \setminus i} \big(1-\phi u_{k\to{}j}^{[\alpha, \beta]}-(1-\phi) z_{k\to{}j}^{[\alpha, \beta]} \big) + 
\bigg[ \prod_{k\in\partial_j^{[\alpha]} \setminus i} \big(1-\phi u^{[\alpha]}_{k\to{}j}-(1-\phi) z^{[\alpha]}_{k\to{}j} \bigg] \times 
(1-\phi)^{k_j^{[2]}} \times
\nonumber\\ 
& \bigg[\prod_{k\in\partial_j^{[\alpha,\beta]} } (1-z_{k\to{}j}^{[\alpha]})-
\prod_{k\in\partial_j^{[\beta]} } (1-z^{[\beta]}_{k\to{}j})
\prod_{k\in\partial_j^{[\alpha,\beta]} } (1-z_{k\to{}j}^{[\alpha,
  \beta]}) \bigg]\Bigg] \; .
\end{align} 

For $z$-type message, the expressions are more complicated.
We have that 
\begin{align}
 \Romannum{1}=
 &\Bigg[ \bigg[ \prod_{k\in\partial_j \setminus i} \big(1-\phi u_{k\to{}j}^{[\alpha]}-(1-\phi) z_{k\to{}j}^{[\alpha]} \big) \bigg]+
(1-\phi)^{k_j^{[1]}-1} \bigg[ 1-\prod_{k\in\partial_j \setminus i}(1-z_{k\to{}j}^{[\alpha]}) \bigg] \Bigg]  \nonumber\\
\Romannum{2}=
 &\Bigg[ \bigg[ \prod_{k\in\partial_j \setminus i} \big(1-\phi u_{k\to{}j}^{[\beta]}-(1-\phi) z_{k\to{}j}^{[\beta]} \big) \bigg]+
(1-\phi)^{k_j^{[2]}-1} \bigg[ 1-\prod_{k\in\partial_j \setminus i}(1-z_{j\to{}i}^{[\beta]}) \bigg] \Bigg]  \nonumber\\
(\Romannum{1},\Romannum{2})=
 &\Bigg[ \bigg[ \prod_{k\in\partial_j \setminus i} \big(1-\phi u_{k\to{}j}^{[\alpha, \beta]}-(1-\phi) z_{k\to{}j}^{[\alpha, \beta]} \big) \bigg]+
(1-\phi)^{k^{[1, 2]}_j-1} \bigg[ 1-\prod_{k\in\partial_j \setminus i}(1-z_{j\to{}i}^{[\alpha, \beta]}) \bigg] \Bigg] +\nonumber\\
& \Bigg[ \bigg[ \prod_{k\in\partial_j^{[\alpha]}} \big(1-\phi u^{[\alpha]}_{k\to{}j}-(1-\phi) z^{[\alpha]}_{k\to{}j} \big) -
(1-\phi)^{k_j^{[\alpha]}}  \prod_{k\in\partial_j^{[\alpha]}}(1-z^{[\alpha]}_{k\to{}j}) \bigg] \times \nonumber\\
& (1-\phi)^{k_j^{[2]}-1} \bigg[ \prod_{k\in\partial_j^{[\alpha,\beta]} \setminus i}(1-z_{k\to{}j}^{[\alpha]}) -
\prod_{k\in\partial_j^{{[\beta]}}}(1-z^{[\beta]}_{k\to{}j}) 
\prod_{k\in\partial_j^{[\alpha,\beta]}\setminus i}(1-z_{k\to{}j}^{[\alpha, \beta]})\bigg]  \Bigg]+ \nonumber\\
& \Bigg[ \bigg[ \prod_{k\in\partial_j^{[\beta]}} \big(1-\phi u^{[\beta]}_{k\to{}j}-(1-\phi) z_{k\to{}j}^{[\beta]} \big) -
(1-\phi)^{k_j^{[\beta]}}  \prod_{k\in\partial_j^{[\beta]}}(1-z ^{[\beta]}_{k\to{}j}) \bigg] \times \nonumber\\
& (1-\phi)^{k_j^{[1]}-1} \bigg[ \prod_{k\in\partial_j^{[\alpha,\beta]} \setminus i}(1-z_{k\to{}j}^{[\beta]}) -
\prod_{k\in\partial_j^{[\alpha]}}(1-z ^{[\alpha]}_{k\to{}j}) 
\prod_{k\in\partial_j^{[\alpha,\beta]} \setminus i}(1-z_{k\to{}j}^{[\alpha, \beta]})\bigg]
\end{align}

\begin{align}
 z_{j\to{}i}^{(1,1),(1,1)} = &1- \Romannum{1} - \Romannum{2} +(\Romannum{1},\Romannum{2}) \nonumber \\
  z_{j\to{}i}^{(1,1),(1,0)} = &\Romannum{2} -(\Romannum{1},\Romannum{2}) \nonumber \\
 z_{j\to{}i}^{(1,1),(0,1)} = &\Romannum{1} - (\Romannum{1},\Romannum{2})
\end{align}

\begin{align}
 z_{j\to{}i}^{(1,0),(1,0)}= 1 -
 &\Bigg[ \bigg[ \prod_{k\in\partial_j \setminus i} \big(1-\phi u_{k\to{}j}^{[\alpha]}-(1-\phi) z_{k\to{}j}^{[\alpha]} \big) \bigg]+
(1-\phi)^{k_j^{[1]}-1} \bigg[ 1-\prod_{k\in\partial_j \setminus i}(1-z_{k\to{}j}^{[\alpha]}) \bigg] \Bigg]  -\nonumber\\
 &\Bigg[ \bigg[ \prod_{k\in\partial_j \setminus i} \big(1-\phi u_{k\to{}j}^{[\beta]}-(1-\phi) z_{k\to{}j}^{[\beta]} \big) \bigg]+
(1-\phi)^{k_j^{[2]}} \bigg[ 1-\prod_{k\in\partial_j \setminus i}(1-z_{j\to{}i}^{[\beta]}) \bigg] \Bigg]  + \nonumber\\
 &\Bigg[ \bigg[ \prod_{k\in\partial_j \setminus i} \big(1-\phi u_{k\to{}j}^{[\alpha, \beta]}-(1-\phi) z_{k\to{}j}^{[\alpha, \beta]} \big) \bigg]+
(1-\phi)^{k^{[1, 2]}_j-1} \bigg[ 1-\prod_{k\in\partial_j \setminus i}(1-z_{j\to{}i}^{[\alpha, \beta]}) \bigg] \Bigg] +\nonumber\\
& \Bigg[ \bigg[ \prod_{k\in\partial_j^{[\alpha]} \setminus i} \big(1-\phi u ^{[\alpha]}_{k\to{}j}-(1-\phi) z_{k\to{}j}^{[\alpha]} \big) -
(1-\phi)^{k_j^{[\alpha]}-1}  \prod_{k\in\partial_j^{[\alpha]} \setminus i}(1-z ^{[\alpha]}_{k\to{}j}) \bigg] \times \nonumber\\
& (1-\phi)^{k_j^{[2]}} \bigg[ \prod_{k\in\partial_j^{[\alpha,\beta]} }(1-z_{k\to{}j}^{[\alpha]}) -
\prod_{k\in\partial_j^{[\beta]}}(1-z ^{[\beta]}_{k\to{}j}) 
\prod_{k\in\partial_j^{[\alpha,\beta]}}(1-z_{k\to{}j}^{[\alpha, \beta]})\bigg]  \Bigg]+ \nonumber\\
& \Bigg[ \bigg[ \prod_{k\in\partial_j^{[\beta]}} \big(1-\phi u ^{[\beta]}_{k\to{}j}-(1-\phi) z_{k\to{}j}^{[\beta]} \big) -
(1-\phi)^{k_j^{[\beta]}}  \prod_{k\in\partial_j^{[\beta]}}(1-z ^{[\beta]}_{k\to{}j}) \bigg] \times \nonumber\\
& (1-\phi)^{k_j^{[1]}-1} \bigg[ \prod_{k\in\partial_j^{[\alpha,\beta]}}(1-z_{k\to{}j}^{[\beta]}) -
\prod_{k\in\partial_j^{[\alpha]} \setminus i}(1-z ^{[\alpha]}_{k\to{}j}) 
\prod_{k\in\partial_j^{[\alpha,\beta]} }(1-z_{k\to{}j}^{[\alpha, \beta]})\bigg]
\end{align}

\begin{align}
 z_{j\to{}i}^{(0,1),(0,1)}= 1 -
 &\Bigg[ \bigg[ \prod_{k\in\partial_j \setminus i} \big(1-\phi u_{k\to{}j}^{[\alpha]}-(1-\phi) z_{k\to{}j}^{[\alpha]} \big) \bigg]+
(1-\phi)^{k_j^{[1]}} \bigg[ 1-\prod_{k\in\partial_j \setminus i}(1-z_{k\to{}j}^{[\alpha]}) \bigg] \Bigg]  -\nonumber\\
 &\Bigg[ \bigg[ \prod_{k\in\partial_j \setminus i} \big(1-\phi u_{k\to{}j}^{[\beta]}-(1-\phi) z_{k\to{}j}^{[\beta]} \big) \bigg]+
(1-\phi)^{k_j^{[2]}-1} \bigg[ 1-\prod_{k\in\partial_j \setminus i}(1-z_{j\to{}i}^{[\beta]}) \bigg] \Bigg]  + \nonumber\\
 &\Bigg[ \bigg[ \prod_{k\in\partial_j \setminus i} \big(1-\phi u_{k\to{}j}^{[\alpha, \beta]}-(1-\phi) z_{k\to{}j}^{[\alpha, \beta]} \big) \bigg]+
(1-\phi)^{k_j^{[1, 2]} -1} \bigg[ 1-\prod_{k\in\partial_j \setminus i}(1-z_{j\to{}i}^{[\alpha, \beta]}) \bigg] \Bigg] +\nonumber\\
& \Bigg[ \bigg[ \prod_{k\in\partial_j^{[\alpha]} } \big(1-\phi u ^{[\alpha]}_{k\to{}j}-(1-\phi) z_{k\to{}j}^{[\alpha]} \big) -
(1-\phi)^{k_j^{[\alpha]}}  \prod_{k\in\partial_j^{[\alpha]} }(1-z ^{[\alpha]}_{k\to{}j}) \bigg] \times \nonumber\\
& (1-\phi)^{k_j^{[2]}-1} \bigg[ \prod_{k\in\partial_j^{[\alpha,\beta]} }(1-z_{k\to{}j}^{[\alpha]}) -
\prod_{k\in\partial_j^{[\beta]}\setminus i}(1-z ^{[\beta]}_{k\to{}j}) 
\prod_{k\in\partial_j^{[\alpha,\beta]}}(1-z_{k\to{}j}^{[\alpha, \beta]})\bigg]  \Bigg]+ \nonumber\\
& \Bigg[ \bigg[ \prod_{k\in\partial_j^{[\beta]}\setminus i} \big(1-\phi u ^{[\beta]}_{k\to{}j}-(1-\phi) z_{k\to{}j}^{[\beta]} \big) -
(1-\phi)^{k_j^{[\beta]}-1}  \prod_{k\in\partial_j^{[\beta]} \setminus i}(1-z ^{[\beta]}_{k\to{}j}) \bigg] \times \nonumber\\
& (1-\phi)^{k_j^{[1]}} \bigg[ \prod_{k\in\partial_j^{[\alpha,\beta]}}(1-z_{k\to{}j}^{[\beta]}) -
\prod_{k\in\partial_j^{{[\alpha]}}}(1-z ^{[\beta]}_{k\to{}j}) 
\prod_{k\in\partial_j^{[\alpha,\beta]} }(1-z_{k\to{}j}^{[\alpha, \beta]})\bigg]
\end{align}


\end{document}